\documentclass[a4paper,11pt]{article}

\usepackage{jcappub}

\usepackage[T1]{fontenc}

\title{\boldmath Some effects of nonlinear vacuum electrodynamics
in strong magnetic and gravitational fields of the pulsar }

\author[a,1]{Medeu Abishev,\note{Corresponding author.}}
\author[a,b]{Yerlan Aimuratov, }
\author[a]{Yermek Aldabergenov}
\author[a]{Nurzada Beissen}
\author[a]{and Meruert Takibayeva }

\affiliation[a]{al-Farabi Kazakh National University,\\ Al-Farabi
av. 71, 050038, Almaty, Kazakhstan } \affiliation[b]{Fesenkov
Astrophysical Institute, \\Observatory 23, 050020, Almaty,
Kazakhstan }

\emailAdd{abishevme@mail.ru}

\abstract{We consider the propagation of X-ray and gamma ray
emissions in strong magnetic and gravitational fields of the
pulsar in nonlinear vacuum electrodynamics. We show that the
radiation has birefrigence.
 We have calculated the
delay between the two modes, as they propagate from the pulsar to
the detecting device.  } 

\begin{document}
\maketitle
\flushbottom

\section{Introduction   \label{sec:intro}}

The field equations in
nonlinear post-Maxwell electrodynamics, which is a direct
consequence of quantum electrodynamics [1]  have the form:
$${1\over \sqrt{-g}}{\partial \over \partial x^\beta}
\Big\{\sqrt{-g}Q^{\sigma\beta}\Big\}= -{4\pi \over c}j^\sigma,
\eqno(1)$$
$$ Q^{\sigma\beta}=8\pi{\partial L\over \partial  F_{\beta\sigma}}
=\Big\{1+\xi\big(\eta_1-2\eta_2\big)I_2\Big\}F^{\sigma\beta}+4\xi\eta_2F^{\sigma\nu}
F_{\nu\mu} F^{\mu\beta},$$
 where $j^\beta$  is the
current density four-vector, $g$ - determinant of the metric
tensor, $\xi=1/B_q^2,$ $I_2=F_{\beta\sigma}F^{\sigma\beta}$
 - invariant of the electromagnetic tensor $F_{\beta\sigma}$ and
according to quantum electrodynamics
$\eta_1=e^2 /(45 \pi \hbar c)=5.1\cdot 10^{-5}, \ \eta_2=7e^2/(180
\pi \hbar c) =9.0\cdot 10^{-5}.$

The second pair of equations of electrodynamics coincides with the
corresponding equations of Maxwell's theory:
$${\partial  F_{\mu\beta} \over \partial x^\nu}+
{\partial  F_{\beta\nu} \over \partial x^\mu} +{\partial
F_{\nu\mu} \over \partial x^\beta}=0.\eqno(2)
$$
Metric tensor in equations (1) satisfy Einstein equations [2]:
$$R_{\beta\sigma}-{1\over 2}g_{\beta\sigma}R
=-{8\pi G\over c^4}T_{\beta\sigma},\eqno(3)$$

where $R_{\beta\sigma}=R^\nu_{\beta\sigma\nu}$ -- Ricci tensor,
$T_{\beta\sigma}$ -- energy-momentum tensor of the matter and all
fields, including electromagnetic. The system of equations (1) -
(3) in our problem will be sought by the method of successive
approximations with a precision linear in the small dimensionless
parameters: the gravitational potential and post-Maxwell
amendments. The gravitational field of the pulsar will be assumed
to be spherically symmetric, and in the harmonic Fock coordinates
[2] metric will be expanded in the small parameter $\alpha/ r$
with the required accuracy:
$$g_{00}=1-{2\alpha\over r},\quad g_{rr}=-1-{2\alpha\over r},\quad
g_{\theta\theta}=r^2g_{rr},\quad
g_{\phi\phi}=g_{\theta\theta}\sin^2\theta,\eqno(4)$$ where
$\alpha =\gamma M/c^2,$ $\gamma$ -- gravitational constant, and
$M$-- mass of the pulsar.

Suppose that at time $t = 0$ from the point ${\bf r} = {\bf r} _0$
of the pulsar magnetosphere hard radiation impulse was emitted.
Then, in magnetic field of the pulsar that impulse, because of
birefringence, will split [3] into two impulses with orthogonal
polarizations and moving at different speeds.

For the convenience of further calculations, we introduce the
spherical coordinate system as follows. Consider a beam of the
first normal mode and draw a tangent to it at the point ${\bf
r}={\bf r}_0.$ Axis of the spherical coordinate system will be
directed in such a way, so that the tangent to the chosen beam and
the center of the pulsar would be lying in the same plane, and
$\theta=\pi/2,$ and the azimuthal coordinate $\phi$ of the source
of hard radiation would be equal to $\phi=0.$

 As it is accepted [4] in the
problems of celestial mechanics, instead of the radial coordinate
$r$ we introduce the coordinate $u = 1/r.$ Then, the non-zero
components of the dipole electromagnetic field tensor of the
pulsar $\bf m$, in the coordinate system $ u,\theta, \phi$, with the
required for our purposes accuracy will be:
$$F_{u\theta }=-F_{\theta u}
=|{\bf m}|\sin\theta_0\sin(\phi-\phi_0),$$
$$F_{u\phi }=-F_{ \phi u}=|{\bf m}|\sin\theta\big[
\sin\theta_0\cos\theta\cos(\phi-\phi_0)-\sin\theta\cos\theta_0\big],\eqno(5)$$
$$ F_{\phi\theta}=-F_{\theta\phi}=2|{\bf m}|u\sin\theta\big[
\sin\theta_0\sin\theta\cos(\phi-\phi_0)+\cos\theta\cos\theta_0\big].$$

According to [3,5]  the
propagation of electromagnetic waves in external electromagnetic
and gravitational fields in nonlinear electrodynamics with field
equations (1) - (3) is equivalent to the propagation of the
normal modes through the  isotropic geodesics in effective
space-time for which metric tensor $G^{eff(1,2)}_{\nu\mu}$ has the
form:
$$G^{eff\ (1,2)}_{\nu\mu}=g_{\nu\mu}-4\eta_{(1,2)}\xi
F_{\nu\beta}g^{\beta\sigma}F_{\sigma\mu }. \eqno(6) $$ Therefore
the study of the laws of propagation of electromagnetic impulses
in magnetic (5) and gravitational (4) fields of a pulsar is
conveniently carried out not by using equations (1) - (2), but
based on the analysis of isotropic geodesics in space-time with
the metric tensor (6).

 Equations for isotropic geodesics in the effective
space-time with the metric tensor (6) will be written in the
form where differentiation is performed not with respect to the
affine parameter $\sigma,$ but with respect to the azimuthal angle
$\phi:$ $${d^2 x^0\over d\phi^2}+\big\{\Gamma^0_{\beta\mu} -{d
x^0\over d\phi}\Gamma^3_{\beta\mu} \big\}{d x^\beta\over d\phi}{d
x^\mu\over d\phi}=0,$$
$${d^2 u\over d\phi^2}+\big\{\Gamma^1_{\beta\mu}
-{d u\over d\phi}\Gamma^3_{\beta\mu} \big\}{d x^\beta\over
d\phi}{d x^\mu\over d\phi}=0,$$
$${d^2 \theta\over d\phi^2}+\big\{\Gamma^2_{\beta\mu}
-{d \theta\over d\phi}\Gamma^3_{\beta\mu} \big\}{d x^\beta\over
d\phi}{d x^\mu\over d\phi}=0,\eqno(7)$$ where
$\Gamma^\nu_{\beta\mu}$ - Christoffel symbols defined in effective
space-time with the metric tensor (6).

The system of equations (7) has a first integral:
$$G^{(1,2)}_{\beta\mu}{dx^\beta\over d\phi}{dx^\mu\over d\phi}=0.
\eqno(8)$$ Equations (7) and (8) are non-linear, but in our
case there are small parameters $\alpha u$ and ${\bf
m}^2\xi\eta_{1,2} u^6.$ Therefore, the solution of these equations
will be sought by the method of successive approximations in these
small parameters.

\section{Solution of the equations for beams }

In the zeroth approximation in small parameters the beam under
mentioned boundary conditions, will be a straight line in the
plane  $\theta=\pi/2,$  passing through the point  $u=u_0,\
\phi=0$ and take a value  $u=u_p$ at pericenter.

Solving this system of
equations, we arrive at the
relations:
 $$u(\phi)=u_p\sin(\phi+\psi),\quad
x^0(\phi)=ct={\cos\psi\over u_p\sin\psi} -{\cos(\phi+\psi)\over
u(\phi)},\quad  \theta(\phi)={\pi\over2}, \eqno(9)$$ where
$\psi$ is defined from: $\sin\psi=u_0/u_p.$

We search the solution of the system of equations (7)-(8) for the
first normal wave in the ordinary form for equations of that type:
$$u(\phi)=u_p\sin(\phi+\psi)+\alpha u_p^2 \Phi_1(\phi)+{\bf m}^2\xi\eta_1
u_p^7\Phi_2(\phi), $$
$$\theta(\phi)={\pi\over 2}+\alpha u_p\Phi_3(\phi)
+{\bf m}^2\xi\eta_1u_p^6\Phi_4(\phi),$$
$$x^0(\phi)={\cos\psi\over u_p\sin\psi}
-{\cos(\phi+\psi)\over u(\phi)}+ \alpha  \Phi_5(\phi)+{\bf
m}^2\xi\eta_1u_p^5\Phi_6(\phi),\eqno(10)$$ where $\Phi_a(\phi),\
a=1-6 $ are unknown functions of the azimuthal angle $\phi,$
having zero order of smallness.

Using an expression (10), we search the solution of the system of
equations (7)-(8) for the first normal wave:$$\Phi_1(\phi)=2+S_1\sin\phi+C_1\cos\phi,\quad
\Phi_2(\phi)={1\over
64}\Big\{f_2(\phi)+S_2\sin\phi+C_2\cos\phi\Big\}, $$
$$\Phi_3(\phi)=S_3\sin\phi+C_3\cos\phi,\quad
\Phi_4(\phi)={\sin2\theta_0\over 64}\Big\{f_4(\phi)
+S_4\sin\phi+C_4\cos\phi\Big\},$$ where
 we use the notation:
$$f_2(\phi)=\sin^2\theta_0\Big\{\cos2(\phi_0+\psi)
\Big[195\phi\cos(\phi+\psi)+65\sin^3(\phi+\psi)+26\sin^5(\phi+\psi)+152\sin^7(\phi+\psi)-$$
$$-144\sin^9(\phi+\psi)\Big]+2\sin2(\phi_0+\psi)
\Big[\big[72\sin^8(\phi+\psi)-40\sin^6(\phi+\psi)
-26\sin^4(\phi+\psi)-$$
$$-39\sin^2(\phi+\psi)\big]\cos(\phi+\psi)
+39\phi\sin(\phi+\psi)\Big]
+32\sin^7(\phi+\psi)-24\sin^5(\phi+\psi)-$$
$$-60\sin^3(\phi+\psi)-180\phi\cos(\phi+\psi) \Big\}-16\Big[2\sin^5(\phi+\psi)+5\sin^3(\phi+\psi)
+15\phi\cos(\phi+\psi)\Big],$$
$$f_4(\phi)=
\Big[75\phi\cos(\phi+\psi)+25\sin^3(\phi+\psi)
+10\sin^5(\phi+\psi)-40\sin^7(\phi+\psi)\Big]\sin(\phi_0+\psi)+$$
$$+\Big[3\phi\sin(\phi+\psi)-[3\sin^2(\phi+\psi)+2\sin^4(\phi+\psi)
+40\sin^6(\phi+\psi)]\cos(\phi+\psi)\Big]\cos(\phi_0+\psi).$$

Solving the equations for functions $\Phi_5(\phi)$ and
$\Phi_6(\phi)$, we have:
$$\Phi_5(\phi)=A_5-2\hbox{ln}\left|
{[1-\cos(\phi+\psi)]\over \sin(\phi+\psi)}\right|,\quad
\Phi_6(\phi)={1\over 64}\Big\{A_6+f_6(\phi)\Big\},$$
where
$$f_6(\phi)=\Big\{16\sin2(\phi_0+\psi)\sin^6(\phi+\psi)
[4-9\sin^2(\phi+\psi)]-\cos2(\phi_0+\psi)\Big[\{144\sin^7(\phi+\psi)+$$
$$+8\sin^5(\phi+\psi) +26\sin^3(\phi+\psi)+39\sin(\phi+\psi)\}\cos(\phi+\psi)-39\phi\Big]
+4[8\sin^5(\phi+\psi)+6\sin^3(\phi+\psi)+$$
$$+9\sin(\phi+\psi)]\cos(\phi+\psi)
-36\phi\Big\}\sin^2\theta_0+8\Big[3+2\sin^2(\phi+\psi)\Big]\sin2(\phi+\psi)
-48\phi.$$
For the constants we have

$$C_1=-2,\quad  S_1=-{2\cos\psi\over(1+\sin\psi)},\quad  C_2=-f_2(0),
$$
$$S_2=f_2(0)\hbox{tg}\psi-{1\over \cos\psi}\Big\{\Big[99\cos2(\phi_0+\psi)
+39(\pi-2\psi)\sin2(\phi_0+\psi)-52\Big]\sin^2\theta_0-112\Big\}.\eqno(11)$$
$$S_3=C_3=0\ \ C_4=-f_4(0),\quad
S_4=5\sin(\phi_0+\psi)\cos\psi\Big[56\sin^6\psi-10\sin^4\psi-$$
$$-15\sin^2\psi-15\Big]+\cos(\phi_0+\psi)\sin\psi\Big[3-280\sin^6\psi+230\sin^4\psi-\sin^2\psi\Big].\eqno(12)$$
$$A_5=2\hbox{ln}\left|{1-\cos\psi\over \sin\psi}\right|,\quad
A_6=-f_6(0)\eqno(13).$$

The beam of the first normal mode after exiting the vicinity of
the pulsar has to be detected by measuring device located in Earth
orbit. Since the nearest pulsars locate [6] at considerable
distance ($r\sim 10$ kps $>> R_n$) from the Earth, it is possible
to assume that in the chosen coordinate system our measuring
device has the coordinate $u_1=1/r_1<<u_p.$ This condition allows
everyone to simply define the required angular coordinates
$\phi_1$ and $\theta_1$ of the device with an aim to register the
beam of the first normal mode. We assume
$\phi_1=\pi-\psi+\beta_1,$ where $\beta_1<<2\pi.$

Substituting this value of  $\phi_1$ in the equation
$u(\phi_1)=u_1,$ and deriving it up to the first order with
respect to $\beta_1$, we will have:
$$\beta_1 =-{u_1\over u_p}+2\alpha u_p\Big[1+{\cos\psi\over(1+\sin\psi)}
\Big]+{{\bf m}^2\xi\eta_1u_p^6\over 64}N_2,$$
$$N_2=S_2\sin\psi-C_2\cos\psi+f_2(\phi=\pi-\psi)=$$
$$={\sin^2\theta_0\over \cos\psi}\Big\{\sin2(\phi_0+\psi)\Big[\{144\sin^8\psi
-80\sin^6\psi-52\sin^4\psi-78\sin^2\psi\}\cos\psi+$$
$$+39(2\psi-\pi)\sin\psi\Big]
+\cos2(\phi_0+\psi)\Big[152\sin^7\psi-144\sin^9\psi+26\sin^5\psi+65\sin^3\psi-99\sin\psi+$$
$$+195(\psi-\pi)\cos\psi
\Big]+4[8\sin^7\psi-6\sin^5\psi-15\sin^3\psi+13\sin\psi+45(\pi-\psi)\cos\psi
]\Big\}+$$
$$+{16\over \cos\psi}\Big[15(\pi-\psi)\cos\psi+7\sin\psi
-5\sin^3\psi-2\sin^5\psi\Big].$$ $$\theta_1=\theta(\phi_1)={{\bf m}^2\xi\eta_1u_p^6\over 64}N_4
\sin2\theta_0,$$
$$N_4=S_4\sin\psi+f_4(\phi=\pi-\psi)+f_4(\phi=0)=
5\sin(\phi_0+\psi)\Big\{\Big[48\sin^7\psi-8\sin^5\psi-$$
$$-10\sin^3\psi-15\sin\psi\Big]\cos\psi+15(\psi-\pi)\Big\}+48\cos(\phi_0+\psi)\Big[4\sin^6\psi-5\sin^8\psi
\Big].$$ This implies that the gravitational field
bends the beams only in one plane.

For the beam on which the pulse propagates carried by the second
normal mode, the expressions (10) take the form:
$$u(\phi)=u_p\sin(\phi+\psi)+\alpha u_p^2\Phi_1(\phi)
+{\bf m}^2\xi\eta_2u_p^7\Phi_2(\phi), $$
$$\theta(\phi)={\pi\over 2}+\alpha u_p\Phi_3(\phi)
+{\bf m}^2\xi\eta_2u_p^6\Phi_4(\phi),$$
$$x^0(\phi)={\cos\psi\over u_p\sin\psi}
-{\cos(\phi+\psi)\over u(\phi)}+ \alpha \Phi_5(\phi) +{\bf
m}^2\xi\eta_2u_p^5\Phi_6(\phi),\eqno(14)$$ with the same
functions $\Phi_a(\phi)$, which were used for the first normal
mode.

Integration constants for beam of the second normal mode are
defined from boundary conditions: at $\phi=0$ and $t=0$ the beam
should begin at the point $u=u_0,\ \theta=\pi/2$ and
asymptotically go to spatial infinity $(r\to\infty,\ u\to 0.)$
Therefore, the integration constants $C_1,\ C_2,\ C_3,\ C_4,\ A_5$
and $A_6$ will be defined by equations (11), (12)-(13).

For the aim of finding values of the integration constants $S_1,\
S_2,\ S_3$ and $S_4,$ we should define the angle $\phi=\phi_2,$ at
which $u=u_1.$ Substituting $\phi=\phi_2=\pi-\psi+\beta_2$ in the
first equation of (14), and equating it to
 $u_1,$ we obtain:
$$\beta_2=-{u_1\over u_p}+\alpha u_p\Big[2+2\cos\psi+S_1\sin\psi)
\Big]+{{\bf m}^2\xi\eta_2u_p^6\over 64}N_{22},$$
$$N_{22}=S_2\sin\psi-C_2\cos\psi+f_2(\phi=\pi-\psi)=S_2\sin\psi+$$
$$+\sin^2\theta_0\Big\{2\sin2(\phi_0+\psi)\Big[112\sin^8\psi
-72\sin^{10}\psi-14\sin^6\psi+13\sin^4\psi-39\sin^2\psi\Big]+$$
$$+\cos2(\phi_0+\psi)\Big[\{152\sin^7\psi-144\sin^9\psi
+26\sin^5\psi+65\sin^3\psi\}\cos\psi+195(\psi-\pi)\Big]
+2\Big[16\sin^7\psi-$$ $$-12\sin^5\psi-30\sin^3\psi\Big]\cos\psi
-180(\psi-\pi)\Big\}-16\sin^3\psi[5+2\sin^2\psi]\cos\psi-240(\psi-\pi).$$

One can use now $\phi=\phi_2=\pi-\psi+\beta_2$ in the second
equation of (7). It is simple to show that
$$\theta_2=\theta(\phi_2)={\pi\over2}+\alpha u_pS_3\sin\psi
+{{\bf m}^2\xi\eta_2u_p^6\over 64}N_{44} \sin2\theta_0,$$
$$N_{44}=S_4\sin\psi+f_4(\phi=\pi-\psi)+f_4(\phi=0)=S_4\sin\psi-5\sin(\phi_0+\psi)\Big\{\Big[8\sin^7\psi-2\sin^5\psi-$$
$$-5\sin^3\psi\Big]\cos\psi-15(\psi-\pi)\Big\}
+\cos(\phi_0+\psi)\Big[40\sin^8\psi-38\sin^6\psi+\sin^4\psi
-3\sin^2\psi\Big].$$

Since at spatial infinity both beams have to get to the measuring
device, the conditions which must be satisfied are the next:
$\beta_1=\beta_2,\ \theta_1=\theta_2.$ Then we
obtain:
 $$S_1=-{2\cos\psi\over(1+\sin\psi)},\quad  S_3=0,$$
$$S_2={u_p\eta_1\over u_0\eta_2}N_2-{u_p\sin^2\theta_0\over u_0}
\Big\{2\sin2(\phi_0+\psi)\Big[112\sin^8\psi
-72\sin^{10}\psi-14\sin^6\psi+$$
$$+13\sin^4\psi-39\sin^2\psi\Big]
+\cos2(\phi_0+\psi)\Big[\{152\sin^7\psi-144\sin^9\psi
+26\sin^5\psi+$$
$$+65\sin^3\psi\}\cos\psi+195(\psi-\pi)\Big]
+2\Big[16\sin^7\psi-12\sin^5\psi-30\sin^3\psi\Big]\cos\psi-$$
$$-180(\psi-\pi)\Big\}+{16u_p\over u_0}\Big\{[5+2\sin^2\psi]\cos\psi\sin^3\psi
-240(\psi-\pi)\Big\},$$
$$S_4={u_p\eta_1\over u_0\eta_2}N_4+{u_p\over u_0}
\Big\{ 5\sin(\phi_0+\psi)\Big[[8\sin^7\psi
-2\sin^5\psi-5\sin^3\psi]\cos\psi-$$
$$-15(\psi-\pi)\Big]-\cos(\phi_0+\psi)\Big[40\sin^8\psi-38\sin^6\psi
+\sin^4\psi-3\sin^2\psi\Big]\Big\}.$$ Thus, all integration
constants for beam of the second mode are defined.

\section{Conclusion}

The last paragraph we define a time interval $T_ {adv} = t_1-t_2,$
which one normal mode ahead of another mode in the propagation of
an electromagnetic pulse from the source to the measurement
device:
$$T_{adv}={{\bf m}^2\xi(\eta_1-\eta_)u_p^5\over 64á}
\Big\{\sin^2\theta_0\Big[16\sin2(\phi_0+\psi)[9\sin^8\psi-4\sin^6\psi]+\cos2(\phi_0+\psi)[\big(144\sin^7\psi+$$
$$+8\sin^5\psi
+26\sin^3\psi+39\sin\psi\big)\cos\psi+39(\pi-\psi)]-4\big(8\sin^5\psi+$$
$$+6\sin^3\psi+9\sin\psi\big)\cos\psi+36(\psi-\pi)\Big]
-16[2\sin^3\psi+3\sin\psi]\cos\psi+48(\psi-\pi)\Big\}.\eqno(15)$$
We estimate the numerical value of $T_ {adv}$ in the case where
the magnetic field source is a neutron star with field on a
surface $B\sim10^{16}$ G (magnetar [7]). Due to the condition
$\xi{\bf m}^2/r^6<<1,$ which must be satisfied for all points of
considered beams, our calculation will be applicable only to the
beams, for which the pericenter $r_p$ exceeds ten radii of the
neutron star. In this spatial region $B(r)<10^{13}$ G and $\xi{\bf
m}^2/r^6\leq 0.05$. Taking into account that the radius of a
typical magnetar is 10 km, from the expression (15) we obtain in
order of magnitude the value: $T_{adv}\sim 10^{-8}$ sec.

\end{document}